\documentclass[twocolumn,english,showpacs,floatfix,preprintnumbers,prb,superscriptaddress]{revtex4}
\usepackage[T1]{fontenc}
\usepackage[latin9]{inputenc}
\usepackage{units}
\usepackage{amsmath}
\usepackage{amssymb}
\usepackage{graphicx}

\makeatletter
\@ifundefined{textcolor}{}
{%
 \definecolor{BLACK}{gray}{0}
 \definecolor{WHITE}{gray}{1}
 \definecolor{RED}{rgb}{1,0,0}
 \definecolor{GREEN}{rgb}{0,1,0}
 \definecolor{BLUE}{rgb}{0,0,1}
 \definecolor{CYAN}{cmyk}{1,0,0,0}
 \definecolor{MAGENTA}{cmyk}{0,1,0,0}
 \definecolor{YELLOW}{cmyk}{0,0,1,0}
 }

\usepackage{float}\usepackage{dcolumn}\usepackage{bm}

\usepackage{babel}

\usepackage{babel}

\usepackage{babel}

\usepackage{babel}

\makeatother

\usepackage{babel}
\begin{document}

\title{Electronic transport within a quasi two-dimensional model for rubrene
single-crystal field effect transistors}

\author{F. Gargiulo}

\email{fernandogargiulo@hotmail.com}

\affiliation{Dipartimento di Scienze Fisiche, Univ. di Napoli {}``Federico II'',
I-80126 Italy}

\author{C.A. Perroni}

\affiliation{Dipartimento di Scienze Fisiche, Univ. di Napoli {}``Federico II'',
I-80126 Italy}

\affiliation{CNR-SPIN}

\author{V. Marigliano Ramaglia}

\affiliation{Dipartimento di Scienze Fisiche, Univ. di Napoli {}``Federico II'',
I-80126 Italy}

\affiliation{CNR-SPIN}

\author{V. Cataudella}

\affiliation{Dipartimento di Scienze Fisiche, Univ. di Napoli {}``Federico II'',
I-80126 Italy}

\affiliation{CNR-SPIN}
\begin{abstract}
Spectral and transport properties of the quasi two-dimensional adiabatic
Su-Schrieffer-Heeger model are studied adjusting the parameters in
order to model rubrene single-crystal field effect transistors with
small but finite density of injected charge carriers. We show that,
with increasing temperature $T$, the chemical potential moves into
the tail of the density of states corresponding to localized states,
but this is not enough to drive the system into an insulating state.
The mobility along different crystallographic directions is calculated
including vertex corrections which give rise to a transport lifetime
one order of magnitude smaller than spectral lifetime of the states
involved in the transport mechanism. With increasing temperature,
the transport properties reach the Ioffe-Regel limit which is ascribed
to less and less appreciable contribution of itinerant states to the
conduction process. The model provides features of the mobility in
close agreement with experiments: right order of magnitude, scaling
as a power law $T^{-\gamma}$, with $\gamma$ close or larger than
two, and correct anisotropy ratio between different in-plane directions.
Due to a realistic high dimensional model, the results are not biased
by uncontrolled approximations. 
\end{abstract}
\maketitle

\section{Introduction}

In recent years, the interest in plastic electronics has grown considerably.
In particular, organic field effect transistors (OFET) are the most
common devices employed to characterize the transport properties of
the organic semiconductors. Single-crystal FET constructed using ultrapure
small molecule semiconductors have allowed to measure mobilities up
to one order of magnitude larger then those typical of thin film transistors.
\cite{takeya,Morpurgo} Among them, rubrene OFETs are much studied.

Transport measurements from low to room temperature in single crystal
OFETs show a behaviour of the charge carrier mobility $\mu$ which
can be defined {}``band-like'' ($\mu\propto T^{-\gamma}$, with
the exponent $\gamma$ close to two) similar to that observed in crystalline
inorganic semiconductors. \cite{Morpurgo} The observation of the
classical Hall effect is also explained in terms of the presence of
relatively free charge carriers, so that the charge conduction is
incompatible with hopping transport between localized states. \cite{Morpurgo}
However, the order of magnitude of mobility is much smaller than that
of pure inorganic semiconductors, and the mean free path for the carriers
has been theoretically estimated to be comparable with the molecular
separation with increasing temperature reaching the Ioffe-Regel limit.
\cite{Cheng}

Extended vs. localized features of charge carriers appear also in
spectroscopic observations. Angle resolved photoemission spectroscopy
(ARPES) supports the extended character of states. \cite{Ostrogota,Ostrogota1}
Indeed, photoemission experiments suggest that the quasi-particle
energy dispersion does exhibit a weak mass renormalization even if
the width of the peaks increases significantly with temperature. On
the other hand, some spectroscopic probes, such as electron paramagnetic
resonance (EPR), \cite{Marumoto1,Marumoto2} THz, \cite{Laarhoven}
and modulated spectroscopy \cite{Sakanoue} are in favor of states
localized within few molecules. The theoretical explanation of the
physical mechanism underlying the conduction properties is still not
fully understood. \cite{Coropceanu}

One of the main problems is to conciliate band-like with localized
features of charge carriers. \cite{Troisi Orlandi} In particular,
the nature of conduction at room temperature with mobilities close
to the Ioffe-Regel limit remains controversial. Actually, in contrast
with other polyacenes, in rubrene, to ascribe the presence of localized
features to small polarons is not likely since the electron-phonon
(el-ph) coupling is not large enough to justify the polaron formation.
\cite{Coropceanu}

A model that is to some extent close to the Su-Schrieffer-Heeger (SSH)
\cite{SSH original} one has been recently introduced. \cite{Troisi Orlandi}
It is a one-dimensional (1D) system where the effect of the electron-phonon
coupling is reduced to a modulation of the transfer integral induced
by low frequency intermolecular modes. This model applies to the most
conductive crystal axis of high mobility systems, such as rubrene.
\cite{Troisi Rubrene} It has been studied by using a dynamic approach
where vibrational modes are treated as classical variables. This approach
has been recently generalized to two dimensions. \cite{Troisi2D}
Within this method, the computed mobility is in agreement with experimental
results. However, the role of dimensionality of the system is not
clear: in fact, in the 1D case, one has $\mu\propto T^{-2}$, while,
in the 2D case, the decrease of the mobility with temperature is intermediate
between $\mu\propto T^{-2}$ and $\mu\propto T^{-1}$. Moreover, this
approach is not satisfying from different points of view. First, the
dynamics of only one charge particle is studied neglecting completely
the role of the chemical potential. Then, the dynamic disorder on
the charge carrier due to the coupling with vibrational modes is included
in a peculiar way: in fact, the electron is not coupled to classical
gaussian fields with time dependent correlation functions which are
determined by the contact of the oscillators with an external bath.
\cite{Madhukar} In this case, the total system is composed only by
one electron (or hole) and vibrational modes, therefore it is not
clear if the corresponding coupled dynamics recovers the thermal equilibrium
on long times and if the Einstein relation can be properly used to
determine the conductivity from the diffusivity. Finally, if the low
frequency modes are not assumed much slower than the electron dynamics
but occur in the same timescale, the quantum nature of phonons cannot
be neglected.

Recently, the transport properties of the 1D SSH model have been analyzed
within a different adiabatic approach. \cite{Ciuchi Fratini} Due
to the geometry of OFET, in which the organic semiconductor is on
the gate, one can assume that the oscillators are at the thermodynamic
equilibrium, therefore the problem is mapped onto that of a single
quantum particle in a random potential (generalized Anderson problem).
\cite{Anderson} However, in order to have finite results of conductivity,
vertex corrections have been completely neglected. Very recently,
some of us have made a systematic study of this 1D model including
the vertex corrections. \cite{Cataudella SSH-1d} While finite frequency
quantities are properly calculated in this 1D model, the inclusion
of vertex corrections leads to a vanishing mobility unless an {\textquotedbl{}ad-hoc\textquotedbl{}}
broadening of the energy eigenvalues is assumed.

In this paper, we generalize a previous work \cite{Perroni Holstein}
in which some of usf have analyzed the properties of electrons coupled
to local modes within the Holstein model by means of the adiabatic
approximation. \cite{Holstein 1,Holstein 2}. We formulate an extension
of the 1D SSH model to the quasi 2D case since this is the relevant
geometry for OFETs geometry. Moreover, the use of a high dimensional
realistic model allows to include the physical anisotropy as an essential
feature and to overcome the difficulties due to the localization of
all the states in reduced dimensions. \cite{Anderson} As discussed
in Appendix, the proper scaling towards the thermodynamic limit is
performed for the quasi 2D models investigated in this paper.

An important characteristic of our study is the small but finite carrier
density with the introduction of the chemical potential. With increasing
temperature, the chemical potential goes towards the tail of localized
states. Therefore, all the experimentally quantities strongly dependent
on the position of the chemical position will probe the features of
localized states. The study of spectral properties clarify that the
states that mainly contribute to the conduction process have low momentum
and are not at the chemical potential. The mobility $\mu$ is studied
as a function of the el-ph coupling and particle density. The behavior
of mobility is in agreement with experiments. Not only the order of
magnitude and the anisotropy ratio between different directions are
right, but also the temperature dependence of $\mu$ is correctly
reproduced since it scales as a power law $T^{-\gamma}$ with $\gamma$
close or larger than two. The inclusion of vertex corrections is relevant
in the calculation, in particular to get a transport lifetime one
order smaller than the spectral lifetime of the states involved in
the transport mechanism. Therefore, high order corrections in two-particle
correlation functions enhance the interaction effects on the properties
of charge carriers. Actually, with increasing temperature, the Ioffe-Regel
limit is reached since the contribution of itinerant states to the
conduction becomes less and less relevant.

The paper is organized as follows. In section II, the model and approach
are discussed in detail. The spectral and transport properties are
discussed in section III and IV, respectively. Finally, section V
contains conclusions and further discussions.

\section{Model and calculation method}

The model studied in this paper include the anisotropy of small molecule
organic semiconductors due to its high dimensionality. The following
Hamiltonian defines the model: 
\begin{equation}
H=\sum_{\mathbf{R}_{i}}\frac{1}{2}ku_{\mathbf{R}_{i}}^{2}+\sum_{\mathbf{R}_{i}}\frac{1}{2}M\dot{u^{2}}_{\mathbf{R}_{i}}+H_{el},\label{eq:hamiltoniana totale}
\end{equation}
 where $u_{\mathbf{R}_{i}}$ is the classical oscillator displacement
at the site in the position $\mathbf{R}_{i}$, $\dot{u}_{\mathbf{R}_{i}}$
the classical oscillator velocity at the site in the position $\mathbf{R}_{i}$,
$M$ the oscillator mass, $k$ the elastic constant. The whole lattice
dynamic is due to a single effective phononic mode whose frequency
$\omega_{0}=\sqrt{K/M}$ is assumed of the order of $5-6meV$. \cite{Troisi Rubrene}

\begin{figure}
\includegraphics[angle=-90,width=0.35\textwidth]{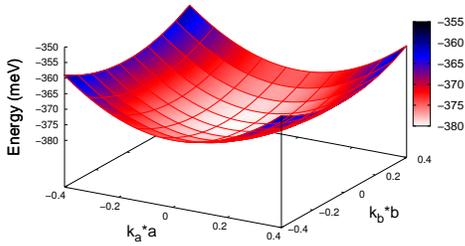}\caption{\label{fig:Dispersion-law} Dispersion law $\varepsilon\left(\mathbf{k}\right)$
for rubrene at $k_{c}=0$.}
\end{figure}

Since we are mostly interested in $dc$ conductivity and low frequency
spectral properties, we need to determine an effective Hamiltonian
for the electron degrees of freedom $H_{el}$ in Eq. (\ref{eq:hamiltoniana totale}
valid for low energy and particle density. We start from the orthorhombic
lattice of rubrene with two molecules per unit cell. \cite{Coropceanu}
The dispersion law of the lowest among the two highest occupied molecular
orbital (HOMO) derived bands is 
\begin{eqnarray}
 &  & \varepsilon\left(\mathbf{k}\right)=-2h_{a}\cos\left(k_{a}a\right)-2h_{b}\cos\left(k_{b}b\right)-2h_{c}\cos\left(k_{c}c\right)\nonumber \\
 &  & -2h_{\frac{a+b}{2}}\cos\left(\frac{k_{a}a+k_{b}b}{2}\right)-2h_{\frac{a-b}{2}}\cos\left(\frac{k_{a}a-k_{b}b}{2}\right)\label{eq:othorimbic dispersion}
\end{eqnarray}
 dropping the site energy diagonal terms. The quantities $a$, $b$,
$c$ represent the lattice parameter lengths along the three crystallographic
vectors of the conventional cell. We assume $a=7.184{\AA}$, $b=14.443{\AA}$,
$c=26.897{\AA}$. \cite{Coropceanu} We follow the experimental paper
by Ding {et al.} in order to extract the transfer integrals. \cite{Ostrogota}
The lowest HOMO derived band is measured through ARPES yielding the
hopping parameters $h_{a}=62.5meV$, $h_{b}=12.5meV$, $h_{\frac{a+b}{2}}=h_{\frac{a-b}{2}}=57.5meV$.
The dispersion law at $k_{c}=0$ is drawn in figure $\left(\ref{fig:Dispersion-law}\right)$.
For small values of $k_{a}$ and $k_{b}$, this graph strictly resembles
a paraboloid. It is clear that an effective dispersion law 
\[
\varepsilon\left(\mathbf{k}\right)=-2t_{a}\cos\left(k_{a}a\right)-2t_{b}\cos\left(k_{b}b\right)
\]
 is capable to fit well the dispersion law (\ref{eq:othorimbic dispersion})
for small values of $k_{a}$ and $k_{b}$ and $k_{c}=0$. Fit estimates
give $t_{a}=118.6meV$ and $t_{b}=68.6meV$. For $t_{c}$ there is
no experimental measure but theoretical estimates seem to agree that
it should be small compared with other directions and it is likely
owing to the large interplanar separation of rubrene. In the following
we assume $t_{c}$ much smaller than $t_{a}$ and $t_{b}$. As we
will discuss later, varying its value seems not too affect much the
calculated mobilities. 

The electronic part $H_{el}$ is then 
\begin{equation}
H_{el}=-\sum_{\mathbf{R}_{i},\vec{\delta}}\bar{t}_{\vec{\delta}}\left(\mathbf{R}_{i}\right)\left[a_{\mathbf{R}_{i}+\vec{\delta}}^{+}a_{\mathbf{R}_{i}}\right].\label{eq:Tight binding SSH}
\end{equation}

\begin{equation}
\bar{t}_{\vec{\delta}}\left(\mathbf{R}_{i}\right)=t_{\vec{\delta}}-\alpha_{\vec{\delta}}\left(u_{\mathbf{R}_{i}}-u_{\mathbf{R}_{i}+\vec{\delta}}\right)\quad\vec{\delta}=\vec{a},\vec{b},\vec{c}
\end{equation}
 In the last formula $a_{\mathbf{R}_{i}}^{+}$ ($a_{\mathbf{R}_{i}}$)
represents the creation (destruction) of a charge carrier in the site
$\mathbf{R}_{i}$, $t_{\vec{\delta}}$ is the bare effective transfer
integral in the direction $\vec{\delta}$, $\alpha_{\vec{\delta}}$
the el-ph parameter that controls the effect on the transfer integral
of the ion displacements in the direction $\vec{\delta}$. Once fixed
$\alpha_{a}$, we impose $\alpha_{b}/\alpha_{a}=t_{b}/t_{a}$ and,
in the same way, $\alpha_{c}/\alpha_{a}=t_{c}/t_{a}$.

The dynamic contained in the term $H_{el}$ is fully quantum, while
we remind that the lattice obeys a classical dynamic. This assumption
is justified by the adiabatic ratio $\hbar\omega/t_{a}\simeq0.05$.
Therefore, the results discussed in this paper are valid starting
from temperatures such that $T>\hbar\omega/k_{B}$, with $k_{B}$
Boltzmann constant. In the adiabatic limit the dimensionless quantity
$\lambda$ 
\begin{equation}
\lambda=\frac{\alpha_{a}^{2}}{4kt_{a}}\label{eq:coupling}
\end{equation}
 is the only relevant parameter to quantify the el-ph coupling strength.
An ab-initio estimate \cite{Troisi Rubrene} of $\lambda_{0}=0.087$
has been given for rubrene and it has been substantially confirmed
later. \cite{Bologna ab-initio,Bologna Raman} This represents our
starting point, but a study of the mobility behaviour when $\lambda$
raises has been performed and is reported in the following sections.
In fact, if $\lambda_{0}$ is claimed to be a good value for the coupling
in one dimension, keeping in mind the increase of the kinetic energy
with dimensionality, the equivalent $\lambda$ should be higher in
three dimensions. Simulations will focus on temperature significantly
lower than $t_{a}/k_{B}$ because this is the typical range investigated
experimentally in OFETs. Moreover, in most OFETs the induced charge
carrier concentration $c=number\: of\: carriers/number\: of\: sites$
is rarely higher than $0.01$ and this is the upper limit for the
concentrations in our simulations. As shown in a previous paper dealing
with a very similar but 1D model, \cite{Cataudella SSH-1d} for this
range of concentrations, the probability distribution of displacements
is only slightly renormalized. Therefore, it is acceptable to refer
to a free-oscillator (Gaussian) distribution for the displacements:
\begin{equation}
P\left(\left\{ u_{\mathbf{R}_{i}}\right\} \right)=\left(\frac{2\pi}{\beta K}\right)^{\frac{L}{2}}\exp\left(-\beta\frac{K}{2}\sum_{i}u_{\mathbf{R}_{i}}^{2}\right)\label{eq:probabili}
\end{equation}
 where $L=L_{a}*L_{b}*L_{c}$ is the finite size of the lattice and
$\beta=\left(k_{B}T\right)^{-1}$. We fix $L_{c}=2$ (i.e. we consider
two crystalline layers) because in OFETs the effective channel of
conductions cover very few planes. \cite{Shehu} Moreover, it is worth
reminding that $t_{c}$ is very small.

We use exact diagonalization to investigate both static and dynamical
properties for the hamiltonian (\ref{eq:hamiltoniana totale}). At
fixed configuration $\left(\left\{ u_{\mathbf{R}_{i}}\right\} \right)$,
one has to diagonalize the Hamiltonian yielding $L$ eigenvalues $E_{r}$,
with $r=1,..,L$. The eigenvector components $U_{i,r}$, with $i=1,..,L$,
are given through the unitary matrix $U$ which diagonalizes the electronic
problem. Each observable results from an average on lattice configurations.
For example, the electron part of the partition function can be calculated
as $\left\langle Z_{el}\left(\left\{ u_{\mathbf{R}_{i}}\right\} \right)\right\rangle $,
where $Z_{el}\left(\left\{ u_{\mathbf{R}_{i}}\right\} \right)$ is
the quantum partition function of the electron subsystem at a given
displacement configuration $\left\{ u_{\mathbf{R}_{i}}\right\} $.
\cite{Z Monte Carlo} Similarly, this would provide approximation-free
evaluation of observables such as mobility and spectral function in
the semiclassical limit. \cite{Observables Monte Carlo} The real
limitation is the accessible finite size of the lattice. The computing
time for each diagonalization is of order $L^{3}$ and this constraints
our analyses up to $L_{a}=L_{b}=24$. In order to reduce the finite
size effect, we use periodic boundary conditions. We have checked
that, for all the static and dynamic quantities studied in this paper,
the thermodynamic limit is reached. The averages are performed by
means of a Monte-Carlo procedure. Actually, we generate a sequence
of random numbers distributed according to $P\left(\left\{ u_{\mathbf{R}_{i}}\right\} \right)$.
For the systems investigated in this paper, to get a good accuracy
even for dynamic quantities we perform the averages on a number of
iterations up to ten thousands.

Within the same framework, it is also possible to calculate the spectral
functions, the density of states, and the conductivity. At fixed configuration,
the density of states is $D(\omega)\left(\left\{ u_{\mathbf{R}_{i}}\right\} \right)$:
\begin{equation}
D(\omega)\left(\left\{ u_{\mathbf{R}_{i}}\right\} \right)=\frac{1}{L}\sum_{i}A_{{\mathbf{R}_{i}},{\mathbf{R}_{i}}}(\omega)\left(\left\{ u_{\mathbf{R}_{i}}\right\} \right),
\end{equation}
 where $A_{{\mathbf{R}_{i}},{\mathbf{R}_{i}}}(\omega)\left(\left\{ u_{\mathbf{R}_{i}}\right\} \right)$
represents the diagonal term of the spectral function 
\begin{equation}
A_{{\mathbf{R}_{i}},{\mathbf{R}_{j}}}(\omega)\left(\left\{ u_{\mathbf{R}_{i}}\right\} \right)=\sum_{r}b_{i,r}^{*}b_{j,r}\delta\left(E_{r}-\omega\right).
\end{equation}
 The spectral function in momentum representation $A_{{\mathbf{k}}}(\omega)$
can be determined after performing the averages over the lattice configurations.
In the case of conductivity, for each configuration, we calculate
the exact Kubo formula \cite{Mahan} 
\begin{eqnarray}
Re\left[\sigma_{\rho,\rho}\left(\omega\right)\left(\left\{ u_{\mathbf{R}_{i}}\right\} \right)\right] & = & \frac{\pi\left(1-e^{-\beta\omega}\right)}{L\omega}\sum_{r\neq s}p_{r}\left(1-p_{s}\right)\nonumber \\
 &  & \left|\left\langle r\left|J_{\rho}\right|s\right\rangle \right|^{2}\delta\left(E_{s}-E_{r}+\omega\right),\label{eq:Kubo formula}
\end{eqnarray}
 where $\rho=a,b,c$, $p_{r}$ is the Fermi distribution 
\begin{equation}
p_{r}=\frac{1}{1+\exp\left(\beta\left(E_{r}-\mu_{p}\right)\right)}\label{eq:Fermi factor}
\end{equation}
 corresponding to the exact eigenvalue $E_{r}$ at any chemical potential
$\mu_{p}$ and $\left\langle r\left|J_{\rho}\right|s\right\rangle $
is the matrix element of the current operator $J_{\rho}$ along the
direction $\hat{e}_{\rho}$, defined as 
\begin{equation}
J_{\rho}=ie\sum_{\vec{R}_{i},\vec{\delta}}\bar{t}_{\vec{\delta}}\left(\mathbf{R}_{i}\right)\left(\vec{\delta}\cdot\hat{e}_{\rho}\right)c_{\vec{R}_{i}}^{\dagger}c_{\vec{R}_{i}+\vec{\delta}},\label{current}
\end{equation}
 with $e$ electron charge. We notice that, in contrast with spectral
properties, the temperature enters the calculation not only through
the displacement distribution, but also directly for each configuration
through the Fermi distributions $p_{r}$. The numerical calculation
of the conductivity is able to include the vertex corrections discarded
by previous approaches.

Finally, the mobility is defined as 
\begin{equation}
\mu_{\rho}=\lim_{\omega\rightarrow0^{+}}\frac{Re\left[\sigma_{\rho,\rho}\left(\omega\right)\right]}{ec}.\label{eq:Mobility}
\end{equation}
 For finite lattice sizes, the delta function appearing in spectral
and transport properties has to be replaced with a Lorentzian, thus
introducing a finite broadening $\eta$ 
\begin{equation}
\delta\left(E+\omega\right)=\lim_{\eta\rightarrow0^{+}}\frac{1}{\pi}\frac{\eta}{\left(E+\omega\right)^{2}+\eta^{2}}.\label{eq:Lorentzian}
\end{equation}
 Clearly, the limit procedure means that the value of $\eta$ should
be kept as small as possible. This does not give any problem for quantities
at finite frequency, as far as $\omega>\eta$. In the case of mobility,
this procedure has to be correctly implemented since there is the
limit $\omega\rightarrow0$. The correct expression is obtained performing
the limits in the following order: $\omega\mapsto0$, $\eta\mapsto0$
and $L\mapsto\infty$. These limits are not trivial, so that in the
Appendix we have reported in detail the {}``calibration procedure''
followed to fix the right $\eta$ for the calculation of the mobility.
In contrast with the 1D case, in our system the mobility dependence
on the broadening $\eta$ exhibits a quasi-plateau with a maximum
whose location gets closer to the zero as the size of the lattice
increases. This makes possible to set $\eta$ as the lowest energy
scale of the system. In addition, it emerges quite clearly that, for
the sizes considered in this work, the system is very close to the
thermodynamic limit. This shows that the broadening $\eta$ is a parameter
of the system which does not influence the mobility estimates. A detailed
analysis has allowed us to obtain values for $\omega_{min}$ and $\eta_{min}$
that assure the determination of the correct value of the mobility
for any temperature. For the mobilities presented in this paper, we
used $\omega_{min}=10^{-3}t_{a}$, $\eta_{min}$ of the order of $2*10^{-2}t_{a}$.

In the following, we will measure energies in units of $t_{a}$. We
will use units such that Boltzmann constant $k_{B}=1$ and Planck
constant $\hbar=1$. We will consider the hopping parameters for rubrene
derived in this section, therefore $t_{b}=0.58$. Along the $c$ axis,
we assume $t_{c}=0.18$. The small effects due to a change in the
value of $t_{c}$ will be discussed in the end of section about transport
properties. Next spectral properties are analyzed.

\section{Spectral properties}

\begin{figure}
\includegraphics[width=0.55\textwidth]{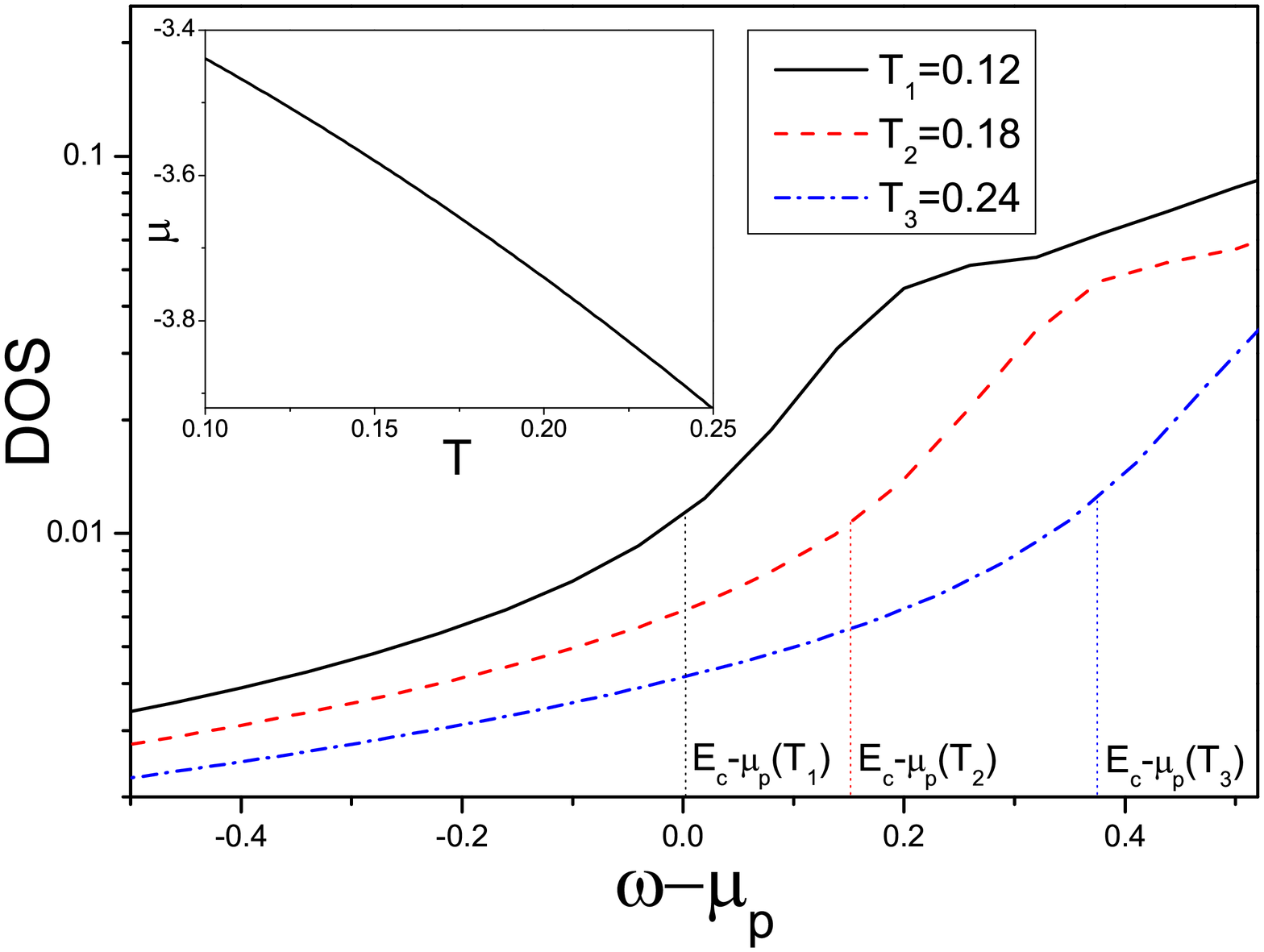}\caption{\label{fig:DOS} The DOS as a function of the frequency (measured
from the chemical potential $\mu_{p}$) at $\lambda=0.12$ and $c=0.002$
for different temperatures. $E_{c}$ is the band edge for free electrons.}
\end{figure}

The study of spectral properties is important to individuate the states
that mainly contribute to the conduction process. In Fig. \ref{fig:DOS},
the Density Of States (DOS) is shown for different temperatures at
$\lambda=0.12$ and $c=0.002$. As shown in logarithmic scale, the
DOS has a tail with a low energy exponential behavior. This region
corresponds to localized states. \cite{economou} We have checked
analyzing the wave functions extracted from exact diagonalizations
that, actually, states with energies deep in the tail are strongly
localized (one or two lattice parameters along the different directions
as localization length). On the other hand, beyond the shoulder (see
Fig. \ref{fig:DOS} for $T_{1}$, $T_{2}$ and $T_{3}$), the itinerant
nature of states is clearly obtained. This analysis gives rough indications
for the mobility edge energy which can be located very close to the
band edge for free electron $E_{c}$ (in our case, $E_{c}=-3.52$).

The states available in the tail increase with temperature. It is
important to analyze the role played by the chemical potential $\mu_{p}$
with varying the temperature. Actually, $\mu_{p}$ enters the energy
tail and will penetrate into it with increasing temperature. At fixed
particle density $c=0.002$, for $T=0.12$, one has $\mu_{p}=-3.49$,
while, for $T=0.24$, $\mu_{p}=-3.88$ (see the inset of Fig. \ref{fig:DOS}).
One important point is that the quantity $E_{c}$ and the close mobility
edge are always over $\mu_{p}$. Therefore, in the regime of low density
relevant for OFET, the itinerant states are not at $\mu$ but at higher
energies. We will show that these states are relevant for the conduction
process. Therefore, the analysis of the properties of a high dimensional
model points out that both localized and itinerant states are present
in the system. This is a clear advantage of our work over previous
studies in low dimensionality \cite{Troisi Orlandi,Troisi2D,Ciuchi Fratini}
in which there are states that are more localized at very low energy
and just less localized close to the free electron edge. Apparently,
in our system, with increasing temperature, more localized states
become available close to $\mu_{p}$, and so the itinerant states
become statistically less effective due to the behavior of the chemical
potential. Eventually, the effect of penetration of $\mu_{p}$ in
the tail will overcome the enlargement of available itinerant states
due to the Fermi statistics. We have checked that the trend of the
chemical potential towards the energy region of the tail is enhanced
with increasing the el-ph coupling $\lambda$. 
\begin{figure}
\includegraphics[width=0.55\textwidth]{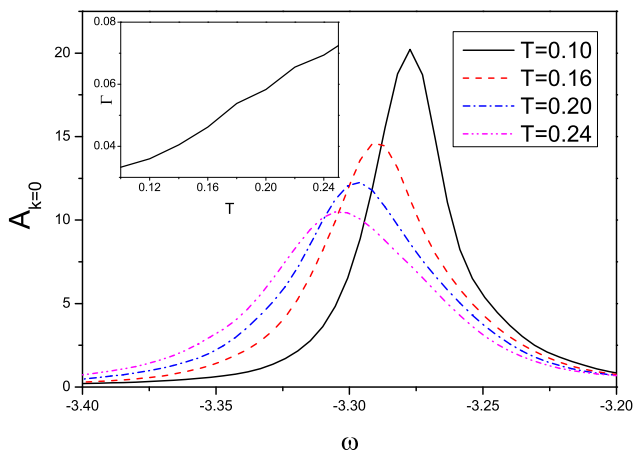}\caption{\label{fig:Spectral} The spectral function at $\mathbf{k}=0$ as
a function of the frequency at $\lambda=0.12$ and $c=0.002$ for
different temperatures. In the inset, the widths at half height of
the spectral function is plotted as function of the temperature.}
\end{figure}

The density of states can be calculated as the sum of the spectral
functions $A_{\mathbf{k}}$ over all the momenta $\mathbf{k}$. We
have checked that the spectral functions with low momentum are more
peaked, while, with increasing $\mathbf{k}$, they tend to broaden.
\cite{Perroni Holstein} The tail in the DOS is due to a marked width
of the high momentum spectral functions. In Fig. \ref{fig:Spectral}
we report the spectral function at $\mathbf{k}=0$ for the same model
parameters as the previous figure. The low damped states close to
$\mathbf{k}=0$ will keep their itinerant character, and they will
be involved into the diffusive conduction process. Actually, we notice
that the spectral function gives a negligible contribution to the
weight of DOS at the chemical potential for all the temperatures.
For example, at $T=0.20$, the spectral weight is concentrated in
an energy region higher than that in which the chemical potential
is located ($\mu_{p}=-3.74$ for $c=0.002$). Actually, the spectral
weight is in the region between $\mu_{p}+2T$ and $\mu_{p}+3T$. Finally,
we point out that, with increasing temperature, the peak position
of the spectral function is only poorly renormalized in comparison
with the bare one, in agreement with results of 1D SSH model. \cite{Cataudella SSH-1d}

It is interesting to estimate the lifetime of the states at low momentum.
The spectral functions at $\mathbf{k}=0$ shift and broaden with increasing
temperature due to the enhanced role of the el-ph coupling. We have
estimated the widths at half height in the inset of Fig. \ref{fig:Spectral}.
At intermediate temperatures, the width is of the order of $0.05t$,
that means of the order of $6$ $meV$. Therefore, the lifetime of
these states is of the order of $22$ $fs$. In the next section,
this lifetime will be compared with the transport lifetime derived
from the mobility.

The analysis of the spectral properties has clarified important features
of the chemical potential and of the states participating at the transport
process. Next, we investigate the {\textquotedbl{}transition rules\textquotedbl{}}
between states involved into the conduction and analyze the mobility
as a function of the particle density and the el-ph coupling.

\section{Transport properties}

\begin{figure}
\includegraphics[width=0.35\textwidth,angle=-90]{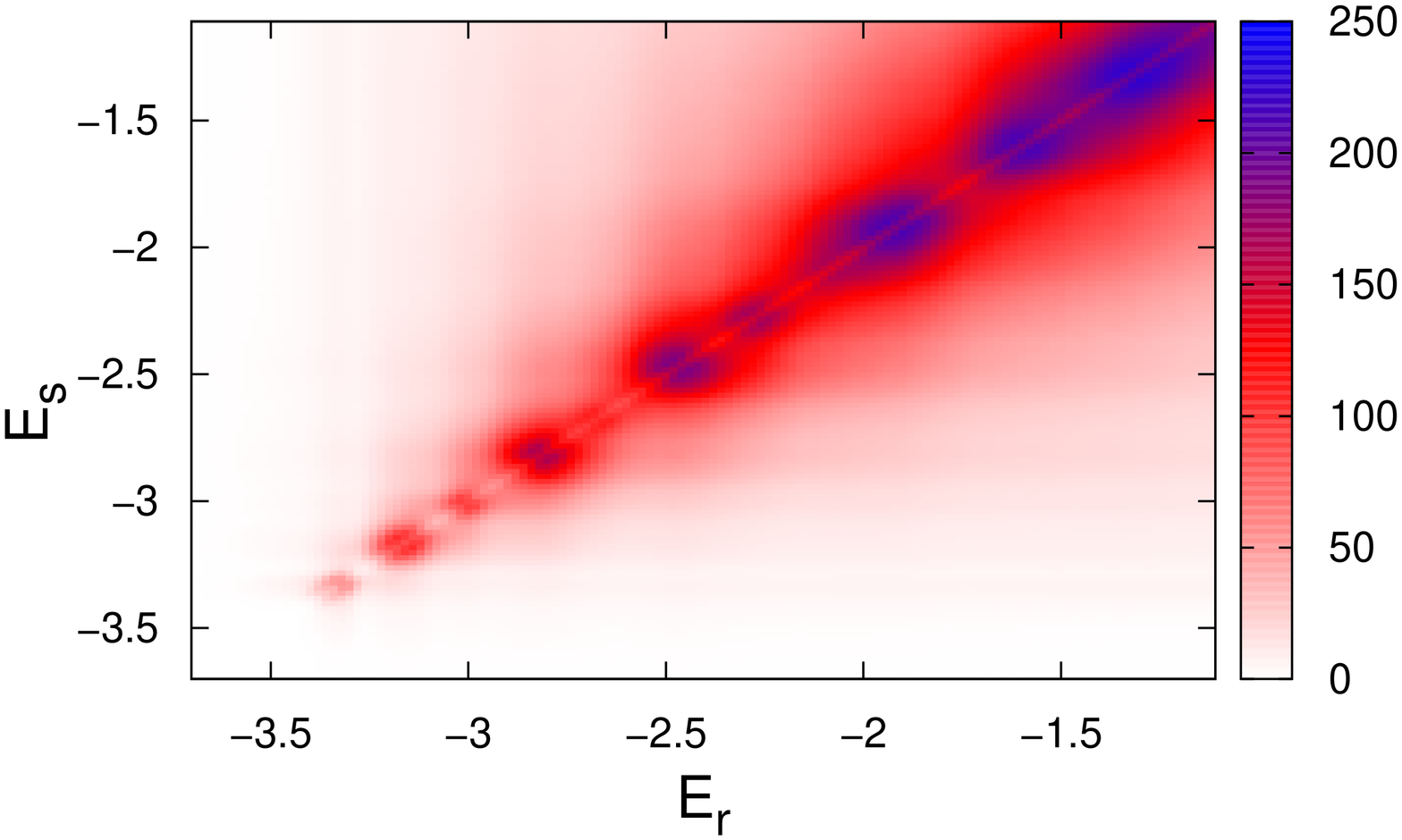} \includegraphics[width=0.35\textwidth,angle=-90]{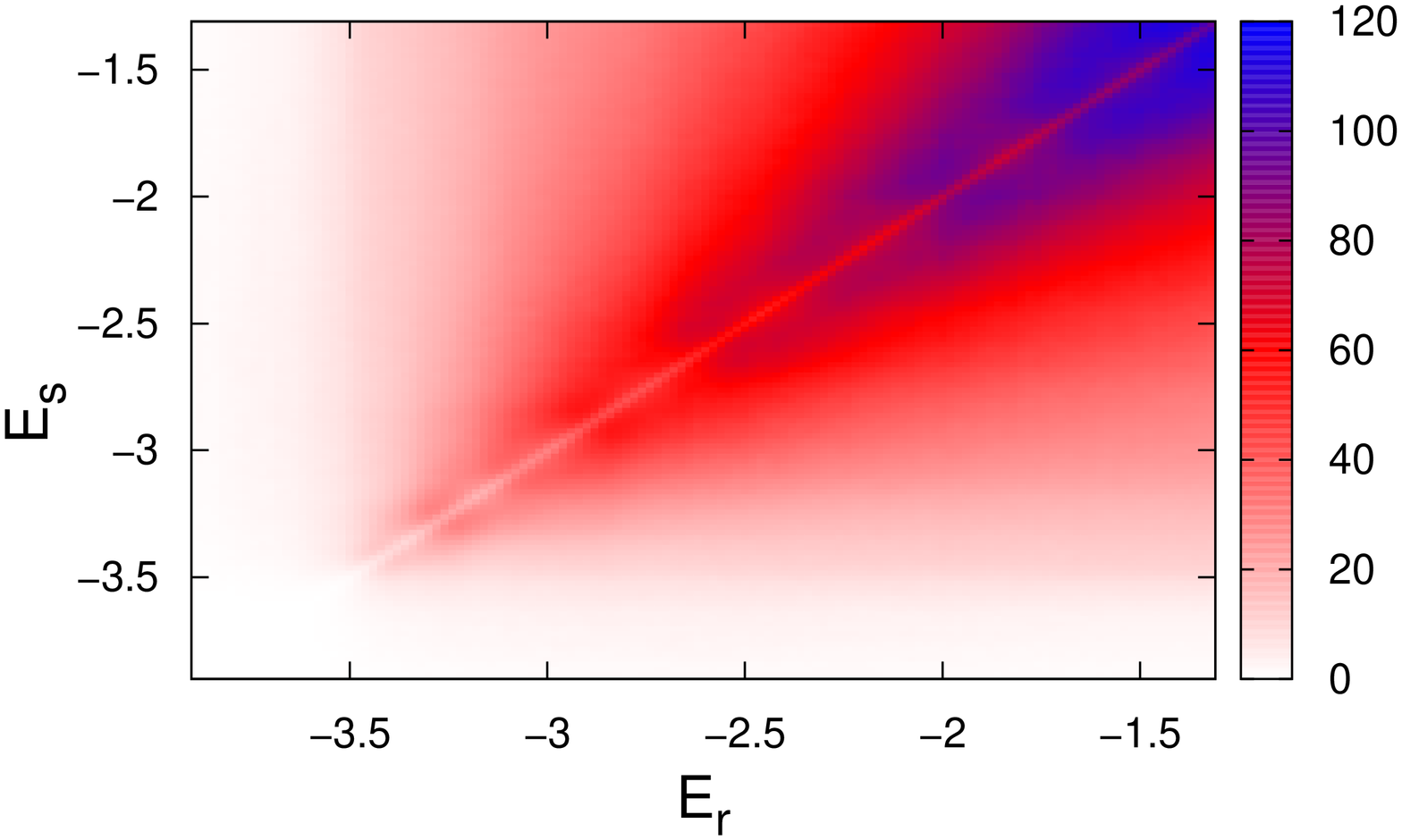}
\caption{\label{fig:Contour} Contour plot of square modulus matrix elements
(in units of $\unitfrac{e^{2}a^{2}}{\hbar^{2}}$) of the current operator
along the direction $a$ between exact eigenstates at $\lambda=0.1$2
and $T_{1}=0.12$ (upper panel), $T_{2}=0.24$ (lower panel) .}
\end{figure}

In this section, the focus will be on the Kubo formula for the conductivity
given in Eq. (\ref{eq:Kubo formula}). We will start from the analysis
of the square modulus matrix elements of the current operator $\left|\left\langle r\left|J_{a}\right|s\right\rangle \right|^{2}$
along the direction $a$ between exact eigenstates $r$ and $s$.
In Fig. \ref{fig:Contour}, we report the contour plot of the current
matrix elements in which the two axes are linked to the energies of
the eigenstates. We consider two different temperatures $T=0.12$
and $T=0.24$. The dark color point towards high values of the matrix
elements, the bright color to low ones. The difference in the intensity
between bright and dark regions correspond to square modulus matrix
elements differing for about two orders of magnitude. The most important
features emerging from the upper plot is that, in the energy range
important for low particle densities, there is a narrow region from
about $-3.4$ to $-3.3$ in which the matrix elements give an appreciable
contribution. Moreover, states close in energy do not repel, but have
sizable overlapping. These features are clearly ascribed to the role
of itinerant states. \cite{economou} We notice that there are several
gaps in the intensities between $-3.3$ and $-2.6$, then, very high
matrix elements occur going toward the center of the electronic band.
At higher temperature (Fig. \ref{fig:Contour}, lower panel), the
contour map is broader, the gaps are almost completely removed and
the absolute magnitude of the matrix elements is lower. We point out
the these {\textquotedbl{}transition rules\textquotedbl{}} are not
strongly dependent on the particle density. We stress that, only for
density $c$ larger than $0.3-0.4$, the matrix elements are sizeable,
and the conduction states are far from the tail of the DOS, therefore
one would get a {\textquotedbl{}metallic state\textquotedbl{}} typical
of an inorganic compound. Indeed, both low density and el-ph interaction
contribute to reduce the conducibilities of OFET in comparison with
inorganic systems. 

\begin{figure}
\includegraphics[width=0.33\textwidth,angle=-90]{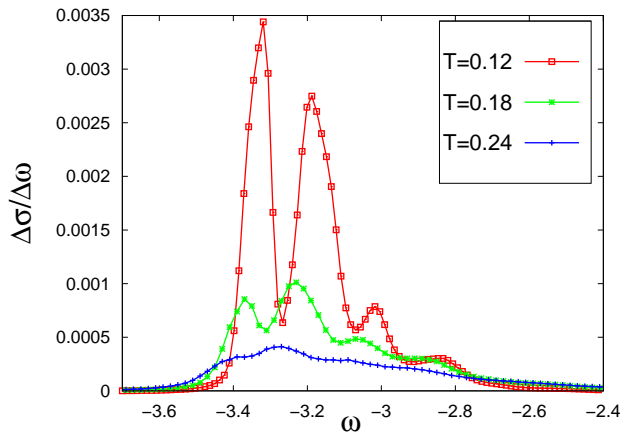}\caption{\label{fig:Resolution} The conducibility contribution along the $a$
direction as function of the energies of the {\textquotedbl{}ingoing
states\textquotedbl{}} at $\lambda=0.12$ and $c=0.002$ for different
temperatures. The unit of the conducibility is $\mu_{0}$ where $\mu_{0}=ea^{2}/\hbar=5.16cm^{2}/(Vs)$.}
\end{figure}

The set of parameters used for the calculation of the mobility are
discussed in Appendix A. Before calculating the mobility as a function
of the temperature, we analyze the mobility along the $a$ direction
as a function of the energy of the states which contribute to it.
More precisely, in Eq. (\ref{eq:Kubo formula}), we sum over the {\textquotedbl{}outgoing
states\textquotedbl{}} $s$ and analyze the results in terms of the
{\textquotedbl{}ingoing states\textquotedbl{}} $r$. We again investigate
the regime of low density relevant for OFET. In Fig. \ref{fig:Resolution},
we plot $\Delta\sigma/\Delta\omega$ as a function of $\omega$. At
low temperature $T=0.12$, the {\textquotedbl{}initial\textquotedbl{}}
states relevant for the conduction are in the energy range where the
current matrix elements are larger, i.e. from about $\omega=-3.4$.
Furthermore, they are in the range between $\mu_{p}+T$ and $\mu_{p}+4T$,
so that they are itinerant. We, finally, note that the curve reproduces
the same gaps as showed in Fig. \ref{fig:Contour}. With increasing
temperature, the itinerant states become more diffusive and distant
from the chemical potential, and, within the energy range close to
$\mu_{p}$, more and more localized states are present. This is the
reason why the mobility gets lower at room temperature where it is
nearly flat as function of the energy. 

\begin{figure}
\includegraphics[width=0.54\textwidth]{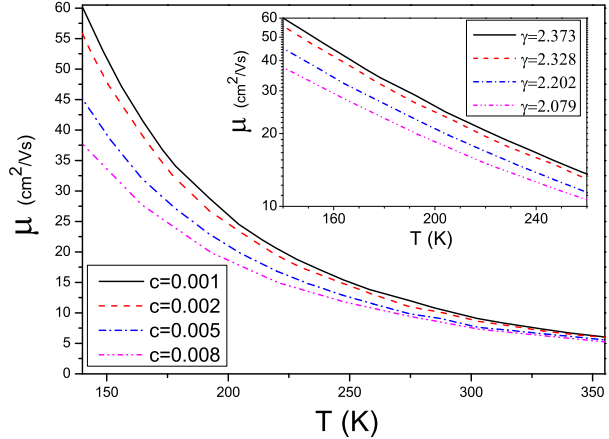} \caption{\label{fig:Mobility lambda fixed} Mobility along the $a$ direction
as a function of the temperature for different values of the concentration
$c$. The coupling strength is fixed at $\lambda=0.1$. The inset
of the lower panel reports the parameters of the fits via the function
$\mu\propto T^{-\gamma}$. }
\end{figure}

In Fig. (\ref{fig:Mobility lambda fixed}), the mobility along the
$a$ direction as a function of the temperature is reported at fixed
coupling $\lambda=0.1$ and for different concentrations $c$. The
upper panel shows that the absolute magnitude of the mobility substantially
agrees with the experimental estimates being $\mu\simeq10cm^{2}/(Vs)$
at room temperature, while, in the inset, where the scale is logarithmic,
the mobility exhibits a quite linear dependence on the temperature
that means a {}``band-like'' behaviour with inverse power-law $\mu\propto T^{-\gamma}$.
The exponent $\gamma$ is evaluated as the slope of the straight lines
drawn in the lower panel and the results that we obtain from the fit
are in the range $2-2.4$, where the highest value is related to the
lower concentration. This trend is in agreement with experimental
measures that for rubrene establish $\gamma\simeq2$ for temperatures
$T>170-180K$. \cite{Morpurgo} A feature of our model is that the
mobility decreases for higher concentrations of carriers. This trend
has been already found in the one-dimensional model. \cite{Cataudella SSH-1d}
Actually, due to the {\textquotedbl{}selection rules\textquotedbl{}}
discussed previously, increasing density can bring the system into
an energy range where the current matrix elements get lower. 

\begin{figure}
\includegraphics[width=0.54\textwidth]{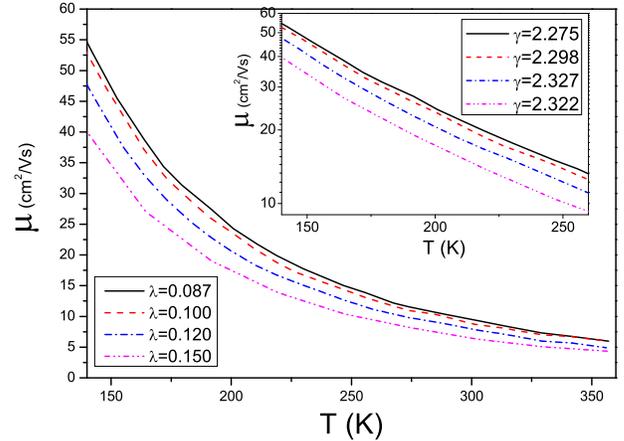} \caption{\label{fig:Mobility concentration fixed} Mobility along the $a$
direction as a function of the temperature for different values of
the coupling strength. The concentration $c$ is fixed at $c=0.002$.
The inset reports the parameters of the fits via the function $\mu\propto T^{-\gamma}$. }
\end{figure}

In Fig. (\ref{fig:Mobility concentration fixed}) the mobility as
a function of the temperature is reported at fixed concentration $c$
and for different couplings starting from the value $\lambda=0.087$.
We just notice two essential features emerging from Fig. (\ref{fig:Mobility concentration fixed}).
Quite obviously, the mobility decreases when the coupling strength
is higher but, differently from the one-dimensional value which predicts
the formation of bond polaron at $\lambda=0.12$, \cite{Cataudella SSH-1d}
in this case this does not occur even at higher values of $\lambda$
for which the {}``band-like'' behaviour still holds. To conclude
this point, we notice that the effect of the coupling on $\gamma$
is less relevant then that of the particle density $c$. 

\begin{figure}
\includegraphics[width=6cm,angle=-90]{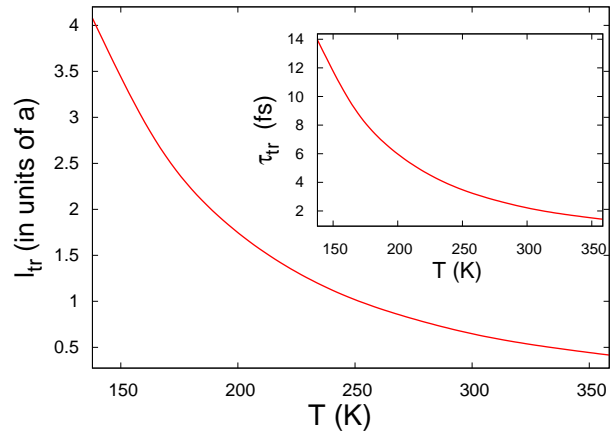} \caption{\label{fig:Scattering} The mean free path (in units of the lattice
parameter $a$) as a function of the temperature at $\lambda=0.12$
and $c=0.002$. The inset reports the scattering time as function
of the temperature for the same model parameters of the main plot}
\end{figure}

Starting from the mobility, we can determine the scattering time from
the relation $\mu=e\tau_{tr}/m$. Since the mass is weakly renormalized
from the el-ph interaction, one can assume $m$ as the bare mass at
$\mathbf{k}=0$. In the inset of Fig. \ref{fig:Scattering}, the scattering
time as a function of the temperature is shown. We point out that
is on the scale of the $fs$, so that it is one order of magnitude
lower than the damping time of the states important for the spectral
properties (on the scale of ten $fs$). Therefore, the transport processes
amplify the effects of the el-ph interaction and the vertex corrections
introduced within our approach are fundamental to take into account
this effect.

From the scattering time, one can deduce the mean free path as $l_{tr}=v_{av}\tau_{tr}$,
where $v_{av}$ is an average velocity of the charge carriers. We
estimate $v_{av}$ from the average kinetic energy as follows: 
\begin{equation}
v_{av}=\sqrt{\frac{2}{m}\left\langle \bar{t}_{\vec{a}}\right\rangle }.
\end{equation}
 The operator $\left\langle \ldots\right\rangle $ has to be intended
both as a thermic average and as an average over the configurations
$\left\{ u_{\mathbf{R}_{i}}\right\} $. The quantity $l_{tr}$ is
reported in Fig. \ref{fig:Scattering}. It is always on the scale
of a few lattice parameters. The most important feature is its temperature
behavior. As a consequence of the el-ph interaction effects, close
to room temperature, it becomes of the order of half lattice parameter
$a$. This means that the Ioffe-Regel limit is reached. \cite{Gunnarsson}
The decrease of the mobility in the Ioffe-Regel limit is not due to
a mass renormalization (dynamic and/or static) but is due to a reduction
of the available itinerant states (the only ones able to transport
current) with the temperature. We remark that this result is due to
the fundamental role played by vertex corrections in the calculation
of the mobility. 

Another important quantity is the anisotropy of the transport properties
along different in-plane directions. Up to now, we have discussed
the mobility only along the $a$ direction, since the corresponding
quantity along the $b$ axis can be roughly reproduced taking into
account the anisotropy factors $t_{b}/t_{a}=0.58$. Actually, the
mobility along $b$ direction roughly scales as the square of the
ratio $t_{b}/t_{a}$, therefore it is about $0.33$ times the mobility
along the $a$ direction. This value is in good agreement with experimental
results. \cite{Morpurgo}

Finally, we briefly discuss the role of hopping parameter $t_{c}$
on the mobility features. Switching from $t_{c}=0.18$ to $t_{c}=0.10$
produces a relative reduction of the mobility that lies between $10\%$
and $15\%$ depending on the concentration of carriers $c$. Although
the sensitivity of the mobility to $t_{c}$ is less relevant than
other parameters, $t_{c}$ can't properly be treated as a {}``dead
parameter''. There may be two reasons for this effect: first reducing
the interplanar coupling yields to a system more bidimensional and
many theoretical predictions state that a purely bidimensional system
should be an insulator, \cite{economou}then, when $t_{c}$ becomes
comparable with the broadening $\eta$ (we keep $\eta$ in the range
$0.1-0.4t_{c}$, see Appendix) the finiteness of the lattice appears
partially invalidating the procedure. Probably, both effects give
their contribution to the reduction of the mobility.

\section{Summary and conclusions}

Spectral and transport properties of the quasi two-dimensional adiabatic
Su-Schrieffer-Heeger model have been discussed with reference to rubrene
single-crystal field effect transistors. An important ingredient of
our model is the small but finite carrier density, therefore the interesting
behavior of the chemical potential as function of temperature has
been investigated. Actually, with increasing temperature, the chemical
potential always enters the tail of the density of states corresponding
to localized states. Therefore, all the experimentally quantities
strongly dependent on the position of the chemical position will probe
the features of localized states. The combined study of spectral and
transport properties is fundamental to shed light on the intricate
conduction process of these materials characterized by small particle
density and intermediate el-ph coupling. From this analysis it emerges
that the states that mainly contribute to the conduction process are
the itinerant ones which are in the energy range between $\mu_{p}+T$
and $\mu_{p}+4T$. With increasing temperature, these states are affected
by enhanced interaction effects and are more distant from $\mu_{p}$,
and, at the same time, more and more localized states become available
in the energy range close to the chemical potential. Therefore, close
to room temperature, the transport properties reach the Ioffe-Regel
limit. Moreover, the transport lifetime is almost one order of magnitude
smaller than the spectral lifetime of the states involved in the transport
mechanism. In order to get this result, it is important to include
vertex corrections into the calculation of the mobility. The mobility
as a function of temperature is in good agreement with experiments:
indeed, the mobility has the right order of magnitude, it scales as
a power law $T^{-\gamma}$, with the $\gamma$ close or larger than
two, and has the correct anisotropy ratio between different in-plane
directions. The use of a realistic quasi 2D model is fundamental for
the analysis of this paper providing reliable results.

The focus of this work has been on the effects of low frequency intermolecular
modes on the transport properties. While the mobility from $100K$
to $300K$ is believed to depend almost entirely by this el-ph coupling,
close to room temperature, the effects of other interactions, for
example that due to local high frequency modes, can give a non negligible
contribution. \cite{Perroni_Comb} Actually, when the mean free path
becomes of the order of the lattice parameter as a result of the only
intermolecular coupling, the local coupling can easily give rise to
an hopping mechanism with an activated mobility. Rubrene is on the
verge of this behavior, that can be accepted as a mechanism for other
poliacenes. As a future work, it could be interesting to use the quasi
2D model discussed in this paper in combination with local el-ph coupling
in order to better understand the conduction mechanism close to room
temperature. Work in this direction is in progress.

\appendix
\begin{figure}
\includegraphics[width=5cm,angle=-90]{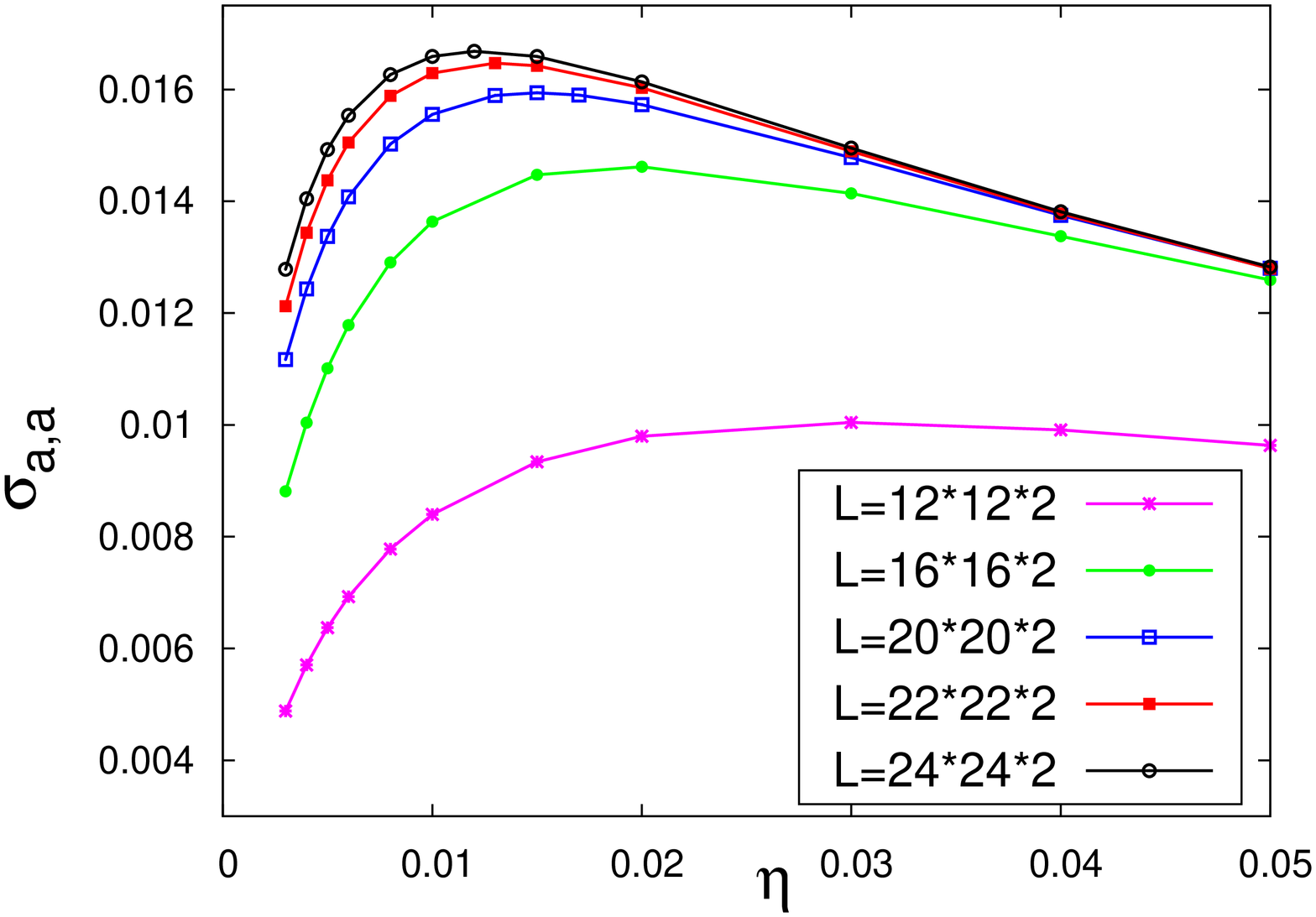} \includegraphics[width=5cm,angle=-90]{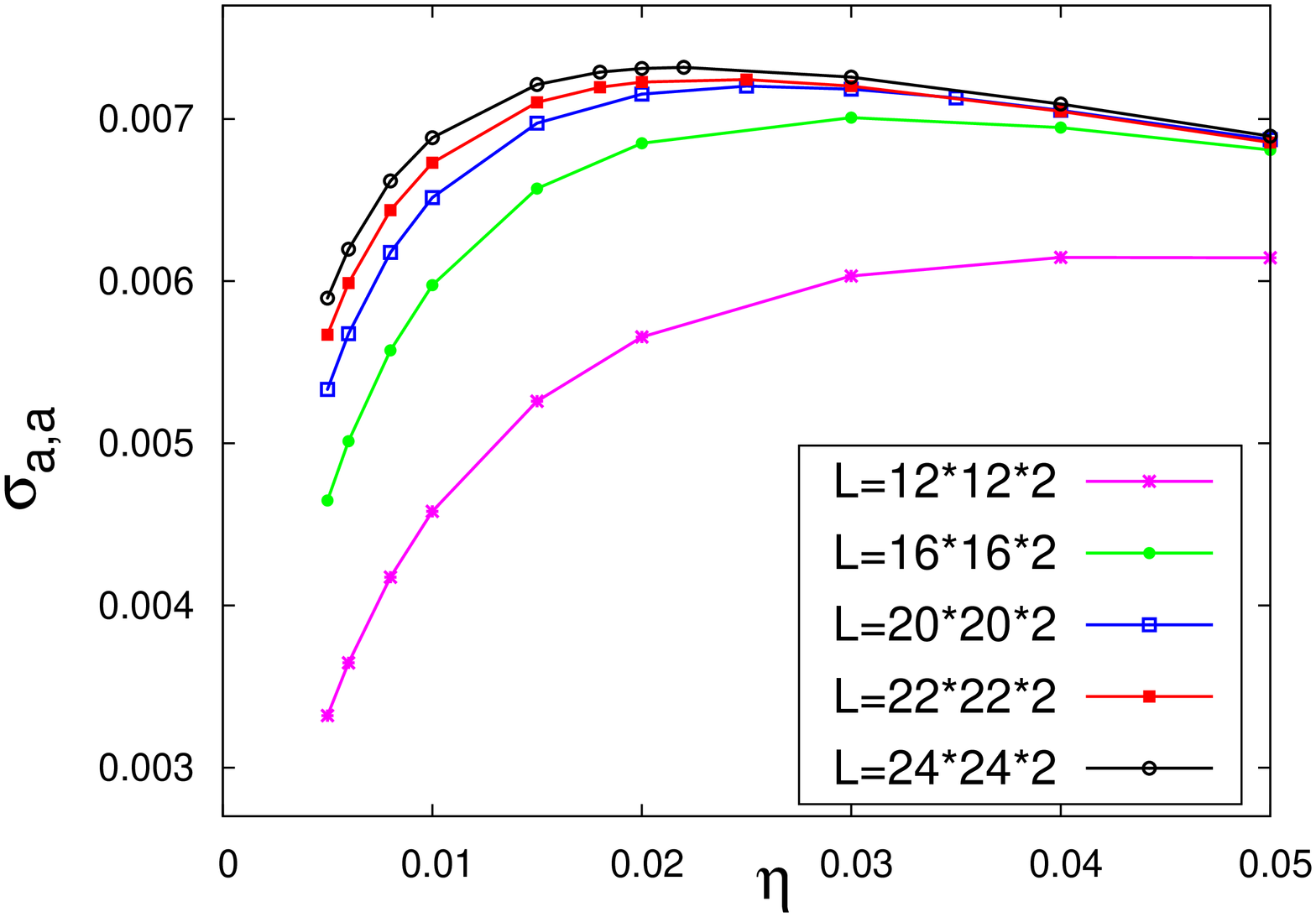}
\includegraphics[width=5cm,angle=-90]{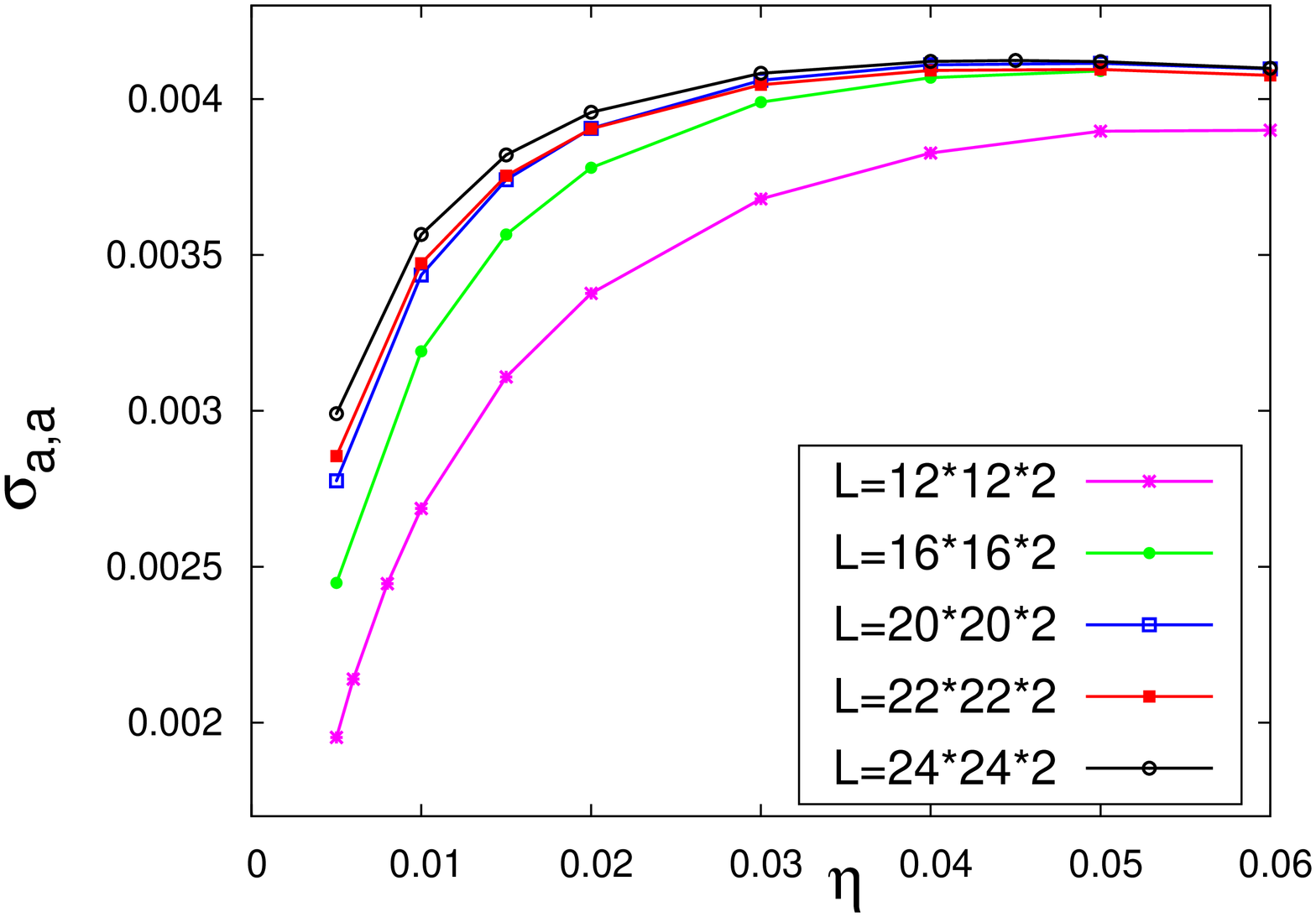} \caption{\label{fig:Calibration} Dependence of the conducibility on $\eta$
for different sizes $L$ of the lattice. Upper panel $T=0.12$. Middle
panel $T=0.18$. Lower panel $T=0.24$. The concentration of carriers
is $c=0.002$ for all the temperatures. The unit of the conducibility
is $\mu_{0}$ where $\mu_{0}=ea^{2}/\hbar=5.16cm^{2}/(Vs)$.}
\end{figure}

\section{{}``Calibration procedure'' }

In this Appendix, we report in detail the {}``calibration procedure''
followed to fix the right value of the broadening $\eta$ for the
calculation of the mobility. According to Eqs. (\ref{eq:Mobility})
and (\ref{eq:Lorentzian}), the mobility is obtained performing the
two limits $\omega\rightarrow0^{+}$ and $\eta\rightarrow0^{+}$ together
with thermodynamic limit $L\rightarrow\infty$. It is clear that the
limit on the size and on $\eta$ has to be done at the same time because,
actually, they are similar. Moreover, it turns out that, for any fixed
choice of the couple $\left(\eta,L\right)$ and for any temperature
$T$, the mobility depends slightly on the frequency $\omega$. There
is a threshold under which the mobility is not sensitive to further
decreasing of $\omega$ that means that we have reached the minimum
frequency that our finite system can resolve. Practically, under the
value $\omega_{0}=10^{-3}$ the mobility shows a plateau. We have
fixed this value for all the calculations of mobility.

In Fig. (\ref{fig:Calibration}), for three different values of temperature,
we have plotted the mobility calculated via the Kubo formula (\ref{eq:Kubo formula})
to analyze how it depends on $\eta$ and the size $L$. The strength
of coupling is fixed at $\lambda=0.1$ and the carrier concentration
is $c=0.002$. At any temperature, the series of data gets closer
with increasing the size of the lattice. Eventually, for the two biggest
sizes, they are not distinguishable for most of the range in which
$\eta$ varies. This allows us to state that, for the maximum size
that we can treat $L_{MAX}=L_{a}*L_{b}*L_{c}=24*24*2$, the system
actually reaches the thermodynamic limit or at least gets very close
to it. For example, in the worst case of low temperature ($T=0.12$
equivalent to $165K$) the two upper curves split only for $\eta<0.02t_{c}$
and it is clear that the value $\eta_{MAX}$ for which the highest
mobility is reached gets closer to zero as the size rises. For all
the regimes of temperature, the mobility exhibits a quasi-plateau
(less definite at low temperatures) and the value of $\eta_{threshold}$
under which the mobility collapses is reduced by increasing the size.
This trend can be easily explained if one notices that, when the size
increases, the separation between two next eigenvalues reduces and
the same happens to the width of the delta function required to couple
a proper number of close eigenvalues. Another feature that is worth
to be noticed is that the curves of different sizes merge together
in better way at high temperatures then at low ones but, at the same
time, for high temperatures $\eta_{MAX}$ is higher. These arguments
lead us to state that, for our model, a correct thermodynamic limit
can be recovered in spite of the analogue 1D model in which the mobility
falls down under a value of $\eta$ that in the limit of infinite
size converges on a finite $\bar{\eta}$. \cite{Cataudella SSH-1d}.
It seems that in 1D the broadening $\eta$ is not just a computational
need that virtually disappears in the infinite size limit but it's
a real missing energy scale without which a finite mobility can't
be obtained. Operatively, for each of the three temperature that appear
in Fig. (\ref{fig:Calibration}), we have determined the value $\eta_{cal}\left(T_{i}\right)$
for which the mobility has a maximum. Then, at any intermediate temperature
$T$, the $\eta_{cal}\left(T\right)$ is obtained from a quadratic
interpolation based on $\eta_{cal}\left(T_{i}\right)$. This procedure
allows to calculate a mobility very close to that one reached in the
thermodynamic limit.


\begin{thebibliography}{References}
\bibitem{takeya}T. Hasegawa and J. Takeya, Sci. Technol. Adv. Mater.
\textbf{10}, 24314 (2009).

\bibitem{Morpurgo} M. E. Gershenson, V. Podzorov and A. F. Morpurgo,
Rev. Mod. Phys. \textbf{78}, 973 (2006).

\bibitem{Cheng}Y. C. Cheng, R. J. Silbey, D. A. da Silva Filho, J.
P. Calbert, J. Cornil, and J. L. Bredas, J. Chem. Phys. \textbf{118}
3764 (2003).

\bibitem{Ostrogota} H. Ding, C. Reese, A. J. Makinen, Z. Bao, and
Y. Gao, Appl. Phys. Lett. \textbf{96}, 222106 (2010).

\bibitem{Ostrogota1} S. I. Machida, Y. Nakayama, S. Duhm, Q. Xin,
A. Funakoshi, N. Ogawa, S. Kera, N. Ueno, and H. Ishii, Phys. Rev.
Lett. \textbf{104}, 156401 (2010).

\bibitem{Marumoto1} K. Marumoto, S. Kuroda, T. Takenobu, and Y. Iwasa,
Phys. Rev. Lett. \textbf{97}, 256603 (2006).

\bibitem{Marumoto2} K. Marumoto, N. Arai, H. Goto, M. Kijima, K.
Murakami, Y. Tominari, J. Takeya, Y. Shimoi, H. Tanaka, S. Kuroda,
T. Kaji, T. Nishikawa, T. Takenobu, and Y. Iwasa, Phys. Rev. B \textbf{83},
075302 (2011).

\bibitem{Laarhoven} H. A. V. Laarhoven, C. F. J. Flipse, M. Koeberg,
M. Bonn, E. Hendry, G. Orlandi, O. D. Jurchescu, T.T.M. Palstra, and
A. Troisi, J. Chem. Phys. \textbf{129}, 044704 (2008).

\bibitem{Sakanoue} T. Sakanoue and H. Sirringhaus, Nat. Mater. \textbf{9}
736 (2010).

\bibitem{Coropceanu} V. Coropceanu, J. Cornil, D. A. da Silva Filho,
Y. Oliver, R. Silbey and J. L. Bredas, Chem. Rev. \textbf{107}, 926
(2007).

\bibitem{Troisi Orlandi}A. Troisi and G. Orlandi, Phys. Rev. Lett.
\textbf{96}, 222106 (2006).

\bibitem{SSH original} W. P. Su, J. R. Schrieffer and A. J. Heeger,
Phys. Rev. Lett\textbf{ 42}, 1698 (1979).

\bibitem{Troisi Rubrene} A. Troisi, Adv. Mat \textbf{19}, 2000 (2007).

\bibitem{Troisi2D} A. Troisi, J. Chem. Phys. \textbf{134}, 034702
(2011).

\bibitem{Madhukar} A. Madhukar and W. Post, Phys. Rev. Lett. \textbf{39},
1424 (1977).

\bibitem{Ciuchi Fratini} S. Fratini and S.Ciuchi, Phys. Rev. Lett.
\textbf{103}, 266601 (2009).

\bibitem{Anderson} P. W. Anderson, Phys. Rev. \textbf{109}, 1492
(1958).

\bibitem{Cataudella SSH-1d} V. Cataudella, G. De Filippis and C.
A. Perroni, Phys. Rev. B \textbf{83}, 165203 (2010).

\bibitem{Perroni Holstein} C. A. Perroni, A. Nocera, V. Marigliano
Ramaglia, and V. Cataudella, Phys. Rev. B \textbf{83}, 245107 (2011).

\bibitem{Holstein 1}T. Holstein, Ann. Phys. \textbf{10} 325 (1959).

\bibitem{Holstein 2}T. Holstein, Ann. Phys. \textbf{10} 343 (1959).

\bibitem{Bologna ab-initio} A. Girlando, L. Grisanti, and M. Masino,
Phys. Rev. B \textbf{82}, 035208 (2010).

\bibitem{Bologna Raman}E. Venuti, I. Billotti, R. G. Della Valle,
A. Brillante, P. Ranzieri, M. Masino and A. Girlando, J. Phys. Chem.
C \textbf{112}, 17416 (2008).

\bibitem{Shehu} A. Shehu, S. D. Quiroga, P. D'Angelo, C. Albonetti,
F. Borgatti, M. Murgia, A. Scorzoni, P. Stoliar, and F. Biscarini,
Phys. Rev. Lett. \textbf{104}, 246602 (2010).

\bibitem{Z Monte Carlo} K. Michielsen and H. de Raedt, Mod. Phys.
Lett. B, \textbf{10}, 467 (1996)

\bibitem{Observables Monte Carlo}E. Dagotto, T. Hotta, A. Moreo,
Phys. Rep. \textbf{344}, 153 (2001).

\bibitem{Mahan} G. D. Mahan, \textit{Many-Particle Physics}, 2nd
ed. (Plenum Press, New York, 1990).

\bibitem{economou} E. N. Economou, \textit{Green's Functions in Quantum
Physics} (Springer Verlag, Berlin, 1983).

\bibitem{Gunnarsson} O. Gunnarson, M. Calandra, and J. E. Han, Rev.
Mod. Phys. \textbf{73}, (2003).

\bibitem{Perroni_Comb} C. A. Perroni, V. Marigliano Ramaglia, and
V. Cataudella, Phys. Rev. B \textbf{84}, 014303 (2011).\end{thebibliography}
\end{document}